\documentclass[aps,prl,twocolumn,showpacs,superscriptaddress,groupedaddress,nofootinbib]{revtex4-2}  
\usepackage[T1]{fontenc}
\usepackage{lmodern} 
\usepackage{graphicx}  
\usepackage{dcolumn}   
\usepackage{bm}        
\usepackage{amssymb}   
\usepackage{amsmath}
\usepackage{bbold}
\usepackage{array}
\usepackage{makecell}
\usepackage{braket}
\usepackage{longtable}
\usepackage{supertabular,booktabs}

\usepackage{titlesec} 
\usepackage[colorlinks,citecolor=blue,urlcolor=blue,hypertexnames=true]{hyperref}
\usepackage{subfigure}

\usepackage[usenames,dvipsnames,svgnames,table]{xcolor} 

\newcommand{\be}{\begin{equation}}
\newcommand{\ee}{\end{equation}}
\newcommand{\bea}{\begin{eqnarray}}
\newcommand{\eea}{\end{eqnarray}}
\newcommand{\bml}{\begin{subequations}}
\newcommand{\eml}{\end{subequations}}
\newcommand{\bfig}{\begin{figure}}
\newcommand{\efig}{\end{figure}}

\newcommand{\bmat}{\begin{pmatrix}}
\newcommand{\emat}{\end{pmatrix}}

\begin{document}

\widetext


\title{\textcolor{Sepia}{\textbf \huge\Large 
 PBH formation in EFT of single field inflation with sharp transition}}

\author{{\large  Sayantan Choudhury${}^{1}$}}
\email{sayantan\_ccsp@sgtuniversity.org,  sayanphysicsisi@gmail.com~(Corresponding author)}
\author{\large Sudhakar~Panda${}^{2,3}$}
\email{panda@niser.ac.in }
\author{ \large M.~Sami${}^{1,4,5}$}
\email{ sami\_ccsp@sgtuniversity.org,  samijamia@gmail.com}

\affiliation{ ${}^{1}$Centre For Cosmology and Science Popularization (CCSP),\\
        SGT University, Gurugram, Delhi- NCR, Haryana- 122505, India,}
\affiliation{${}^{2}$School of Physical Sciences,  National Institute of Science Education and Research, Bhubaneswar, Odisha - 752050, India,}
\affiliation{${}^{3}$ Homi Bhabha National Institute, Training School Complex, Anushakti Nagar, Mumbai - 400085, India,}
\affiliation{${}^{4}$Center for Theoretical Physics, Eurasian National University, Astana 010008, Kazakhstan.}
	\affiliation{${}^{5}$Chinese Academy of Sciences,52 Sanlihe Rd, Xicheng District, Beijing.}

\begin{abstract}
Using the Effective Field Theory (EFT) framework of single field inflation, we investigate the possibility of the formation of Primordial Black Holes (PBHs) in the Slow Roll (SR) to Ultra Slow Roll (USR) {\it sharp} transition.  We demonstrate that, due to one-loop correction to the power spectrum, causality is violated ($c_s>1$) for the mass range of PBHs, $M_{\rm PBH}>10^{2}{\rm gm}$ created during the said transition. We find that non-canonical features with $c_s<1$ worsen the predictions of the canonical framework of single-field inflation.



\end{abstract}

\maketitle
In the framework under consideration, the basic idea is to begin with an effective action that is model independent and valid below the Ultra Violet (UV) cut-off scale. Additionally, the structure of the Effective Field Theory (EFT) action is constrained by symmetry \cite{Weinberg:2008hq,Cheung:2007st,Choudhury:2017glj}. 
This  framework allows us to provide a constraint in terms of the effective speed of sound $c_s$, which can be expressed in terms EFT parameters. We adopt the unitary gauge 
while working with
St$\ddot{u}$ckelberg technique, which essentially includes
scalar perturbation dubbed Goldstone modes.
In this case, the corresponding scalar perturbation variable, $\delta\phi$, is eaten up by the graviton (metric), which has then three physical degrees of freedom: spin-$0$ scalar perturbation itself and the other two are represented by two-component spin-$2$ tensor helicity modes.
This phenomenon exactly mimics the spontaneous symmetry breaking in the $SU(N)$ non-abelian gauge theory. 
To produce a massive spin $1$ degrees of freedom in the unitary gauge, the associated gauge boson eats up the non-linearly transformed goldstone mode under the implementation of the underlying gauge symmetry. Just like the Standard Model Higgs sector, one can think of embedding the Goldstone mode within the framework of non-linear sigma model which can further be interpreted as the UV completed version of the linearized gauge symmetry.

In general, the framework under consideration, might include several scalar degrees of freedom, however,  we have limited our discussion to a single  scalar field in EFT. The advantage of this formalism is that it deals with both the canonical and non-canonical scalar fields 
and allows us to simultaneously investigate PBH formation in a general setting,
see Refs \cite{Hawking:1974rv,Carr:1974nx,Carr:1975qj,Chapline:1975ojl,Carr:1993aq,Choudhury:2013woa}
and \cite{Kristiano:2022maq,Riotto:2023hoz,Choudhury:2023vuj,Choudhury:2023rks,Choudhury:2023hvf,Choudhury:2023kdb,Kristiano:2023scm,Riotto:2023gpm,Firouzjahi:2023ahg,Firouzjahi:2023aum,Franciolini:2023lgy,Cheng:2023ikq,Tasinato:2023ukp,Choudhury:2023hvf,Choudhury:2023kdb,Motohashi:2023syh}. 


Let us  now consider the following diffeomorphic transformation,
\bea	
	&&t\longrightarrow t+ \xi^{0}(t,{\bf x}),~x^{i}\longrightarrow x^{i}~~~\forall~ i=1,2,3\nonumber\\
 &&\delta\phi\longrightarrow\delta\phi +\dot{\phi}_{0}(t)\xi^{0}(t,{\bf x}).\quad\quad
\eea
The temporal diffeomorphism parameter is given by, $\xi^{0}(t,{\bf x})$. In this case, we pick up the gravitational gauge so that $\phi(t,{\bf x})=\phi_{0}(t),$ where $\phi_{0}(t)$ is the background time-dependent scalar field in homogeneous isotropic FLRW cosmic space-time.
Additionally, this necessitates that in this gauge choice, 
$\delta \phi(t,{\bf x})=0~.$

In the present context we start with the following abbreviated EFT action \cite{Cheung:2007st,Choudhury:2017glj}:
\begin{widetext}
\bea
 S&=&\displaystyle\int d^{4}x \sqrt{-g}\left[\frac{M^2_{pl}}{2}R+M^2_{pl} \dot{H} g^{00}-M^2_{pl} \left(3H^2+\dot{H}\right)+\frac{M^{4}_{2}(t)}{2!}\left(g^{00}+1\right)^2+\frac{M^{4}_{3}(t)}{3!}\left(g^{00}+1\right)^3~~~~~~~~\nonumber\right.\\&&
	\left.\displaystyle~~~~~~~~~~~~~~~~~~~~~~~~\quad\quad\quad\quad\quad\quad\quad\quad\quad\quad-\frac{\bar{M}^{3}_{1}(t)}{2}\left(g^{00}+1\right)\delta K^{\mu}_{\mu}-\frac{\bar{M}^{2}_{2}(t)}{2}(\delta K^{\mu}_{\mu})^2-\frac{\bar{M}^{2}_{3}(t)}{2}\delta K^{\mu}_{\nu}\delta K^{\nu}_{\mu}\right].
	\eea
 \end{widetext}
 Here $\delta K_{\mu\nu}=\left(K_{\mu\nu}-a^2Hh_{\mu\nu}\right),$  where $K_{\mu\nu}$ is the extrinsic curvature at constant time, $a$ the scale factor and $H$ represents Hubble parameter in quasi de-Sitter space, $h_{\mu\nu}$ is the spin-$2$ field;  $M_{2}(t)$, \textcolor{red}{$M_{3}(t)$}, $\bar{M}_{1}(t)$, $\bar{M}_{2}(t)$ and $\bar{M}_{3}(t)$ represent the Wilson coefficients which slowly changes with the time scale for this EFT set up. 
 

The Goldstone mode ($\pi(t, {\bf x})$) transforms as follows under the time diffeomorphism symmetry, 
$\pi(t, {\bf x})\rightarrow\tilde{\pi}(t,{\bf x})=\pi(t, {\bf x})-\xi^{0}(t,{\bf x}),$
   where the local parameter is represented by $\xi^{0}(t,{\bf x})$. These Goldstone modes serve as an analogue for the scalar modes' function in cosmic perturbation in this study. Next, the the condition for fixing the unitary gauge is given by, 
  $\pi(t,{\bf x})=0$ which implies $\tilde{\pi}(t,{\bf x})=-\xi^{0}(t,{\bf x})~.$
   We must now comprehend the decoupling limit in greater depth in order to build the EFT action. The gravity and Goldstone modes' mixing contributions in this limit are easily disregarded. Let's start with the EFT operator $-\dot{H}M_{pl}^2g^{00}$, which is necessary for further computation to demonstrate the veracity of this assertion.
The temporal part of the metric can be expressed as follows, 
$g^{00}=-1+\delta g^{00},$
where $\delta g^{00}$ denotes the perturbation. The remaining contributions are a kinetic contribution $M_{pl}^2\dot{H}\dot{\pi}^2\bar{g^{00}}$ and a mixing contribution $M_{pl}^2\dot{H}\dot{\pi}\delta g^{00}$.
A canonical normalised metric perturbation is also used, $\delta g^{00}_c=M_{pl}\delta g^{00}$,  and mixing contribution is given by,  $M_{pl}^2\dot{H}\dot{\pi}\delta g^{00}= \sqrt{\dot{H}}\dot{\pi}_c\delta g^{00}_c$.  In the decoupling limit, one may conveniently ignore the mixing term above the energy scale $E_{mix}=\sqrt{\dot{H}}$. Another option is to combine contributions $M_{pl}^2\dot{H}\dot{\pi}^2\delta{g^{00}}$ and $\pi M_{pl}^2\ddot{H}\dot{\pi}\bar{g}^{00}$, which may be represented as follows after canonical normalization,  
$M_{pl}^2\dot{H}\dot{\pi}^2\delta{g^{00}}=\dot{\pi}_c^2\delta{g^{00}_c}/M_{pl},\pi M_{pl}^2\ddot{H}\dot{\pi}\bar{g}^{00}=\ddot{H}\pi_c\dot{\pi}_c\bar{g}^{00}/\dot{H}.$ The contribution from the $M_{pl}^2\dot{H}\dot{\pi}\delta{g^{00}}$ term can be disregarded for $E>E_{mix}$. 


In the decoupling limit,  the second-order Goldstone EFT action ($S^{(2)}_{\pi}$) is given by:
  	 \bea 
  	S^{(2)}_{\pi}&=&\displaystyle \int d^{4}x ~a^3\left(\frac{-M^2_{pl}\dot{H}}{c^2_s}\right)\left(\dot{\pi}^2-c^2_s\frac{\left(\partial_{i}\pi\right)^2}{a^2}\right).~~~~~\quad\quad\eea 
 where the effective sound speed for the present EFT set up is defined as,
   $c_{s}\equiv \left[1-\frac{2M^4_2}{\dot{H}M^2_p}\right]^{-1/2}.$
   In the current context, the spatial components of the metric fluctuation is provided by:
   \bea g_{ij}\sim a^{2}(t)\left[\left(1+2\zeta(t,{\bf x})\right)\delta_{ij}\right]~~\forall~~~i=1,2,3,\eea
   where the scalar fluctuation is denoted by $\zeta(t,{\bf x})$, where $\zeta(t,{\bf x})\approx-H\pi(t,{\bf x})~.$ Further,  in terms of comoving curvature perturbation the second-order EFT action ($S^{(2)}_{\zeta}$) can be re-expressed as:
   \bea S^{(2)}_{\zeta}&=&M^2_{pl}\displaystyle \int d\tau\;  d^3x\;  a^2\;\left(\frac{\epsilon}{c^2_s}\right)\left(\zeta^{'2}-c^2_s\left(\partial_i\zeta\right)^2\right).~~~~~\quad\quad\eea  
   A new variable, $v=\frac{a\sqrt{2\epsilon}}{c_s}M_{ pl}\zeta$ with $\epsilon=1-{\cal H}^{'}/{\cal H}^2$, also referred to as the Mukhanov Sasaki (MS) variable, is then introduced. \textcolor{red}{Throughout the paper, we denote the derivative with respect to conformal time by a symbol $'$.} After that, using the Fourier transformation, and varying the action, the MS equation for the scalar modes can be expressed as:
   \bea
 v^{''}_{\bf k}+\left(c^2_sk^2-\frac{2}{\tau^2}\right)v_{\bf k}=0\,.
   \eea
The following expression provides the solution of the Mukhanov Sasaki equation for the scalar mode during the SR period ($\tau< \tau_s$) using Bunch Davies initial condition:
   \bea
 \zeta_{\bf k}(\tau)&=&\left(\frac{ic_sH}{2M_{\rm pl}\sqrt{\epsilon}}\right)_*\frac{\left(1+ikc_s\tau\right)}{(c_sk)^{3/2}}\; e^{-ikc_s\tau}.
 \eea
Here all the terms appearing in the parenthesis bracket are evaluated at the pivot scale $k_*=0.02{\rm Mpc}^{-1}$. For this reason, we identify the conformal time-dependent effective sound speed at this point as, $c_s(\tau_*)=c_{s}$, which will appear frequently in the rest of the computations performed in this paper. In principle, $c_s$ can be less than, greater than, or equal to unity, depending upon the 
 EFT setup of interest
 Let us note that, during the SR phase, the first slow-roll parameter, $\epsilon$ can be treated to be a constant. The another slow-roll parameter in this discussion, $\eta=\epsilon^{'}/\epsilon {\cal H}$,  is extremely small and less than unity during the SR phase. In the rest of the computation of the paper both of these slow-roll parameters, particularly the behaviour of $\eta$ plays a significant role. 
 
In this work, we have considered  a sharp transition from SR to an ultra-slow-roll (USR) phase. 
 The USR phase  persists for a very small span and at the end of the USR phase inflation ends. In terms of the conformal time scale, SR phase is valid just up to $\tau_s$ where the mentioned sharp transition occurs. The choice of the phases is based on our previous work \cite{Choudhury:2023vuj}, where we have found that to have a sufficient number of e-foldings, which is $\Delta {\cal N}_{\rm Total}=\Delta{\cal N}_{\rm SR}+\Delta{\cal N}_{\rm USR}\sim {\cal O}(55-60)$, it is justifiable to have only these mentioned two consecutive phases,  SR and USR respectively. Such a choice is completely motivated to incorporate one-loop effects in the primordial power spectrum for scalar modes and to generate small mass PBHs, $M_{\rm PBH}\sim 10^2{\rm gm}$ at the momentum scale $k_{\rm PBH}=k_s\sim 10^{21}{\rm Mpc}^{-1}$. We have also found from our analysis in ref. \cite{Choudhury:2023rks} that due to strong constraints  coming from renormalization and ressumation of the computed power spectrum, it is not a good option to push the PBH formation scale to $k_{\rm PBH}=k_s\sim 10^{5}{\rm Mpc}^{-1}$ corresponding to large mass PBHs, $M_{\rm PBH}\sim 10^{31}{\rm kg}\sim M_{\odot}$, where $M_{\odot}$ represents the solar mass.  We also found that adding an extra SR phase after the completion of the mentioned USR phase does not help. The prime reason for this conclusion is because of the fact that having the first slow-roll phase (SRI), then a sharp transition from SRI to short-lived USR, then another sharp transition from USR to second slow-roll phase (SRII) followed by successful inflation in the presence of renormalization and resummation  can not be achieved. We did the analysis very carefully with canonical single field inflationary paradigm in ref \cite{Choudhury:2023vuj} and found that in this specific case, the total number of e-foldings is achieved, $\Delta {\cal N}_{\rm Total}\sim {\cal O}(20-25)$ with SRI, USR, and SRII to achieve large mass PBHs at $k_{\rm PBH}=k_s\sim 10^{5}{\rm Mpc}^{-1}$. Later to understand this situation from a more general perspective and to provide a more strong argument regarding the validity of the assumption of choosing the sequence of the mentioned phases, in ref. \cite{Choudhury:2023rks} we did the comparison between cases: (1) SR to USR, and (2) SRI to USR and to SRII in the presence of EFT framework based on all classes of $P(X,\phi)$ theories where sharp transition phenomena are implemented. Interestingly, We have found from the analysis performed in ref. \cite{Choudhury:2023rks} that if renormalization and resummation of the power spectrum is done correctly then with the addition of a new SRII phase,  successful inflation can not be accomplished  if we demand the generation of large mass PBHs at the scale $k_{\rm PBH}=k_s\sim 10^{5}{\rm Mpc}^{-1}$. On the other hand, if we demand the completely opposite i.e. the generation of extremely small mass PBHs from the EFT setup at the scale, $k_{\rm PBH}=k_s\sim 10^{21}{\rm Mpc}^{-1}$, then from both the scenarios (SR+USR and SRI+USR+SRII with sharp transitions) we get the same result. In this case, the total number of e-foldings achieved from the setup will be, $\Delta {\cal N}_{\rm Total}\sim {\cal O}(55-60)$ from both the scenarios. For this reason, it is enough to consider the scenario having only SR and USR phases followed by the end of inflation. The main reason for the confusion regarding this issue comes from the tree-level computation of the primordial power spectrum for the scalar modes. At the tree level, to generate large mass PBHs at the scale $k_{\rm PBH}=k_s\sim 10^{5}{\rm Mpc}^{-1}$, an additional SRII phase is necessarily required to achieve a total number of e-foldings, $\Delta {\cal N}_{\rm Total}\sim {\cal O}(55-60)$. But the renormalization and ressumation on the one-loop corrected power spectrum spoils the tree level result; we have found that at least for the sharp transition it is not possible to generate large mass PBHs at the scale, $k_{\rm PBH}=k_s\sim 10^{5}{\rm Mpc}^{-1}$. Keeping the aforesaid in mind, we have restricted our analysis to SR and USR phases, where the transition from the SR to USR phase is occurring sharply.   
 
 Before proceeding ahead, let us mention the findings from other studies that have been performed to address the same issue from a different perspective. 
 For completeness, let's compare our results with the findings of other authors.
  Very recently, in ref. \cite{Riotto:2023gpm,Firouzjahi:2023ahg,Firouzjahi:2023aum} the authors have pointed out that with the help of a smooth transition from SRI to USR and USR to SRII, it is possible to suppress the contribution from large amplitude fluctuation as appearing in the one-loop corrected expression for the primordial power spectrum for the scalar modes. In these studies, it is also pointed out that it is possible to generate large mass PBHs at the scale, $k_{\rm PBH}=k_s\sim 10^{5}{\rm Mpc}^{-1}$ which can accommodate the sufficient number of e-foldings necessarily required to validate inflation. Some more studies have been done along the same subject line in the refs. \cite{Franciolini:2023lgy,Cheng:2023ikq,Tasinato:2023ukp,Choudhury:2023hvf,Choudhury:2023kdb,Motohashi:2023syh}. In particular, a sudden or smooth transition drastically changes the final result on the estimated PBH mass and in the corresponding PBH abundance. For sharp transitions, the one-loop corrected power spectrum is almost unaffected without taking into account the renormalization \cite{Riotto:2023gpm,Firouzjahi:2023ahg,Firouzjahi:2023aum,Franciolini:2023lgy,Cheng:2023ikq,Tasinato:2023ukp,Choudhury:2023hvf,Choudhury:2023kdb,Motohashi:2023syh}, and with renormalization and resummation extremely sensitive and not allowing large mass PBHs generation from the underlying setup \cite{Choudhury:2023vuj,Choudhury:2023rks}. On the other hand, for the smooth transition in refs. \cite{Riotto:2023gpm,Firouzjahi:2023ahg,Firouzjahi:2023aum} the authors have explicitly shown that the final result of the one-loop corrected power spectrum is strongly suppressed. However, none of the studies with smooth transition \cite{Riotto:2023gpm,Firouzjahi:2023ahg,Firouzjahi:2023aum} and some of the studies with sharp transition \cite{Kristiano:2022maq,Riotto:2023hoz,Kristiano:2023scm,Franciolini:2023lgy,Cheng:2023ikq,Motohashi:2023syh} have addressed the crucial issue of renormalization and resummation to finally arrive at the conclusion regarding the suppression of one-loop contributions as well as the generation of large mass PBHs. 
  In the next half of the paper, with the detailed computation, we will establish all of the presented arguments in support of our analysis.

The USR region, which is visible within the window $\tau_s\leq \tau\leq \tau_e$, will be added to our discussion in the paragraphs to follow.  Here, $\tau_ s$ and $\tau_e$ are the conformal time scales of the sharp transition from the SR to the USR and at the end of inflation, respectively. The conformal time dependency of the first slow-roll parameter in the USR regime can be expressed as, 
$\epsilon(\tau)=\epsilon  \;\left(\tau/\tau_s\right)^{6}.$  Here $\epsilon$ is the slow-roll parameter in the SR region which is approximately constant and less than unity to validate slow-roll approximation.

The following is the formula for the curvature perturbation resulting from the solution of the MS equation in the USR region:
 \bea
 &&\zeta_{\bf k}(\tau)=\left(\frac{ic_sH}{2M_{ pl}\sqrt{\epsilon}}\right)\left(\frac{\tau_s}{\tau}\right)^{3}\frac{1}{(c_sk)^{3/2}}\nonumber\\
 &&\left[\alpha_{\bf k}\left(1+ikc_s\tau\right)\; e^{-ikc_s\tau}-\beta_{\bf k}\left(1-ikc_s\tau\right)\; e^{ikc_s\tau}\right],\quad\quad\quad
 \eea
Thus the solutions for the USR and SR regions are actually connected by the Bogoliubov coefficients $\alpha_{\bf k}$ and $\beta_{\bf k}$.  Since the initial vacuum is chosen to be Bunch Davies, these coefficients are fixed by using the continuity of the modes and its associated momenta at the SR to USR sharp transition at the time scale $\tau=\tau_s$. Utilizing these facts we found that: 
\bea \alpha_{\bf k}&=&1-\frac{3}{2ik^{3}c^{3}_s\tau^{3}_s}\left(1+k^{2}c^{2}_s\tau^{2}_s\right),\nonumber\\
\beta_{\bf k}&=&-\frac{3}{2ik^{3}c^{3}_s\tau^{3}_s}\left(1+ikc_s\tau_s\right)^{2}\; e^{-2ikc_s\tau_s}.\eea
Here it is important to note that, we must explicitly quantize the appropriate scalar modes in order to compute the formula for the two-point correlation function and the related power spectrum in Fourier space, which is needed to calculate the cosmological correlations.

Let's now take our analysis a step further and directly compute the impact of the power spectrum's one-loop correction from the perturbation's scalar modes using the curvature perturbation, which expands the typical EFT action in third order:
\bea &&\label{s3} S^{(3)}_{\zeta}=\int d\tau\;  d^3x\;  M^2_{ pl}a^2\; \bigg(\left(3\left(c^2_s-1\right)\epsilon+\epsilon^2-\frac{1}{2}\epsilon^3\right)\zeta^{'2}\zeta\nonumber\\
&&\quad\quad\quad\quad\quad\quad+\frac{\epsilon}{c^2_s}\bigg(\epsilon-2s+1-c^2_s\bigg)\left(\partial_i\zeta\right)^2\zeta\nonumber\\ 
&&\quad\quad\quad\quad\quad\quad-\frac{2\epsilon}{c^2_s}\zeta^{'}\left(\partial_i\zeta\right)\left(\partial_i\partial^{-2}\left(\frac{\epsilon\zeta^{'}}{c^2_s}\right)\right)\nonumber\\
&&\quad\quad\quad\quad\quad\quad-\frac{1}{aH}\left(1-\frac{1}{c^2_{s}}\right)\epsilon \bigg(\zeta^{'3}+\zeta^{'}(\partial_{i}\zeta)^2\bigg)
    \nonumber\\
&&
   \nonumber\\
&&\quad\quad\quad\quad\quad\quad +\frac{1}{2}\epsilon\zeta\left(\partial_i\partial_j\partial^{-2}\left(\frac{\epsilon\zeta^{'}}{c^2_s}\right)\right)^2
\nonumber\\
&&\quad\quad\quad\quad\quad\quad+\frac{1}{2c^2_s}\epsilon\partial_{\tau}\left(\frac{\eta}{c^2_s}\right)\zeta^{'}\zeta^{2}+\cdots\bigg),\quad\quad\eea

     \begin{figure}[htb!]
    	\centering
 \subfigure[For $c_s=0.6(<1)$  with $M^4_2/\dot{H}M^2_p\sim -0.89$ (non-canonical and causal).]{
      	\includegraphics[width=8.7cm,height=5.3cm] {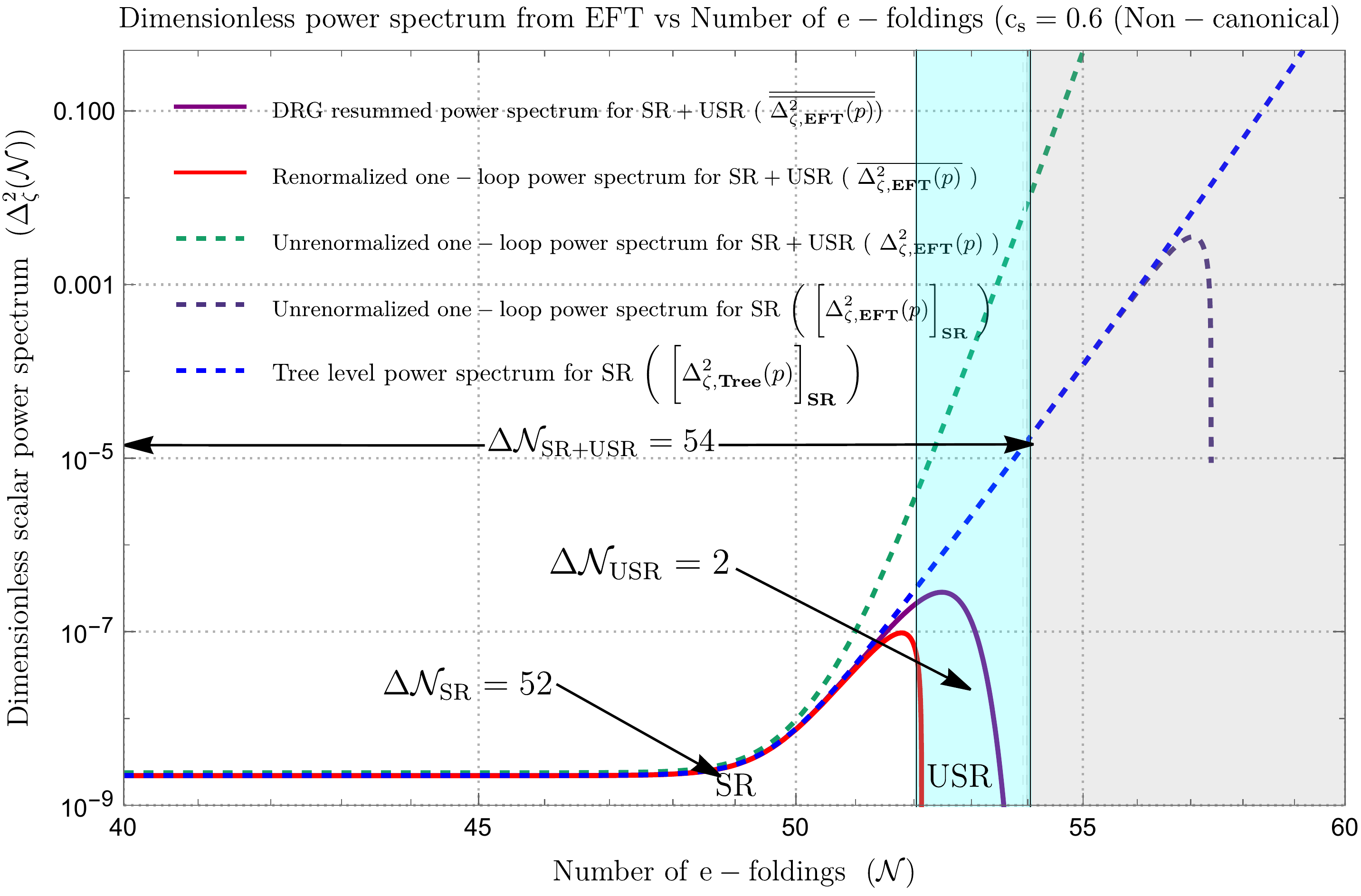}
        \label{N1}
    }
    \subfigure[For $c_s=1$  with $M^4_2/\dot{H}M^2_p\sim 0$ (canonical and causal).]{
       \includegraphics[width=8.7cm,height=5.3cm] {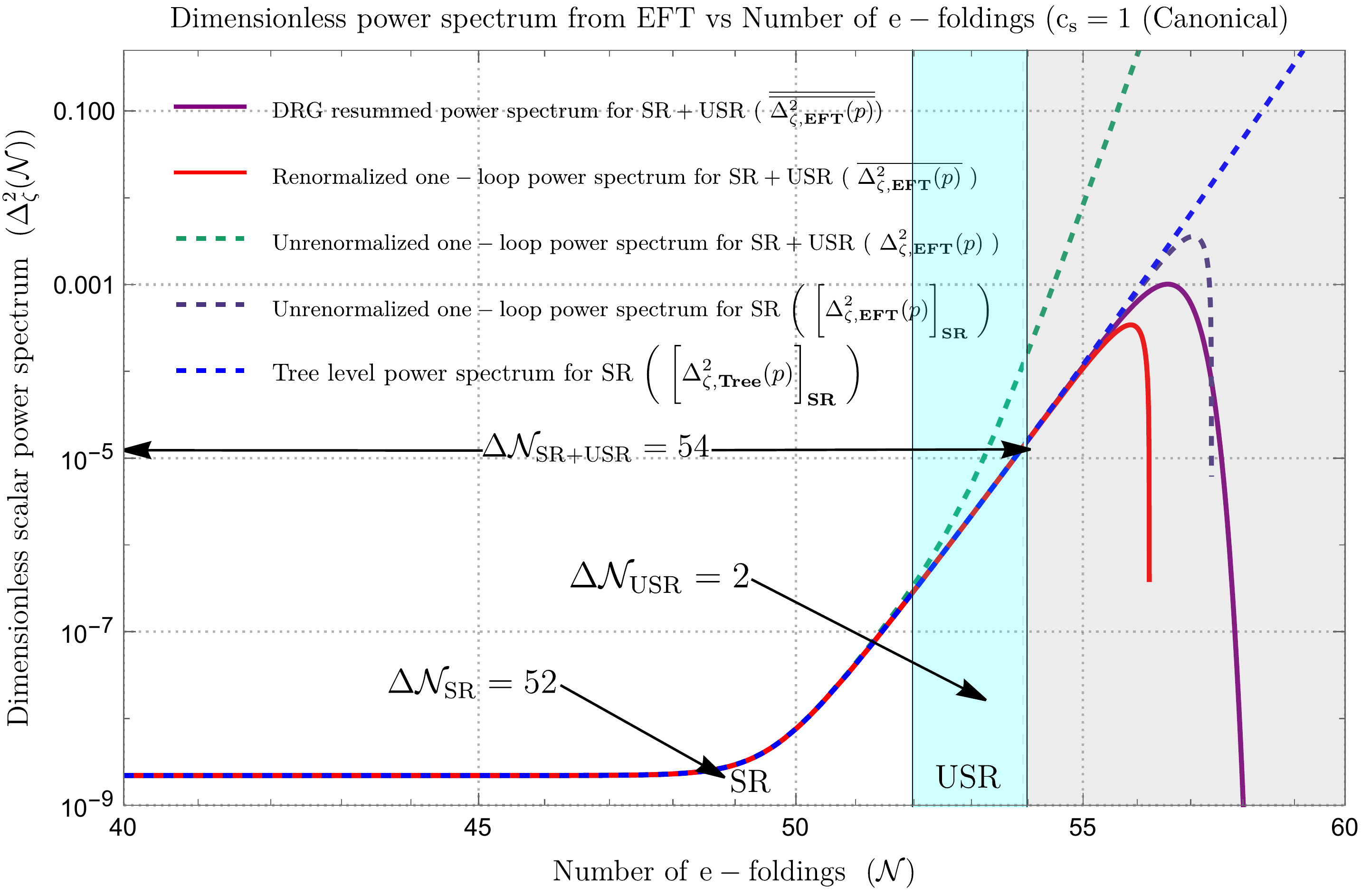}
        \label{N2}
       }
        \subfigure[For $c_s=1.17(>1)$  with $M^4_2/\dot{H}M^2_p\sim 0.13$ (non-canonical and a-causal).]{
       \includegraphics[width=8.7cm,height=5.3cm] {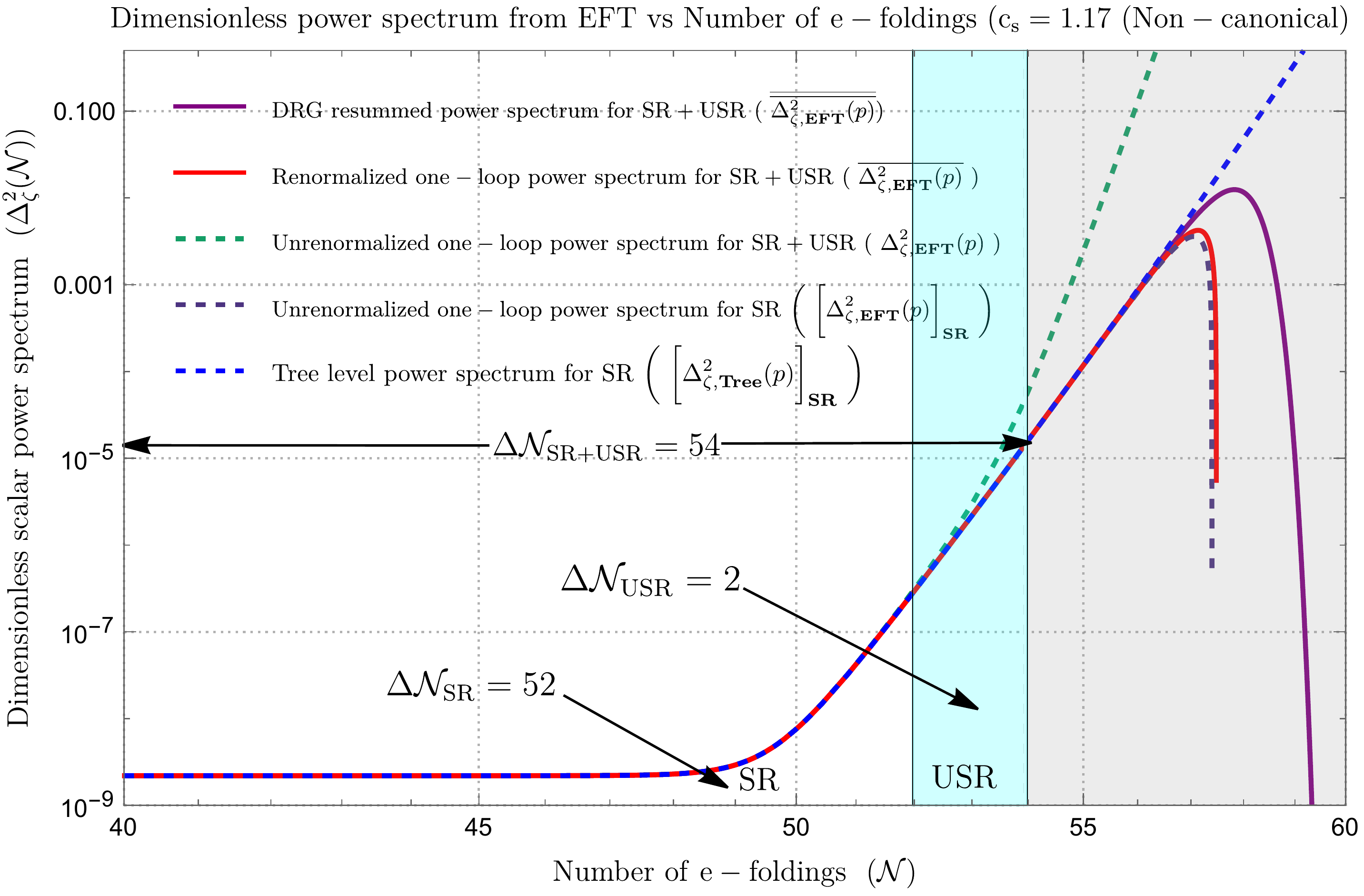}
        \label{N3}
       }
    	\caption[Optional caption for list of figures]{Behaviour of the dimensionless power spectrum for scalar modes with respect to the number of e-foldings. Plots show that $c_s\gtrsim 1$ is the allowed range of effective sound speed. } 
    	\label{Spectrum3}
    \end{figure}
Here $\eta =\epsilon^{'}/\epsilon {\cal H}$, $s=c^{'}_s/aHc_s$ and $\cdots$ represent contributions which are highly suppressed. Due to the appearance of the parameter $s$ it is also clear that the effective sound speed parameter is not constant through the cosmological evolution. We will mention the underlying behaviour and the corresponding parametrization of the effective sound speed in the two consecutive phases, SR, USR, during SR to USR sharp transition and at the end of inflation respectively.  It is important to note that the above-mentioned third order action for the scalar perturbation can be derived by explicit use of the well-known ADM formalism. For the detailed computation of the above-mentioned action for $c_s\neq 1$, see refs. \cite{Chen:2006nt,Chen:2010xka,Chen:2009zp}, where the authors have explicitly derived this in detail utilizing the ADM formalism. For the special case, where $c_s=1$ in the case of canonical single field models, the equivalent derivation of the third-order action can be found in refs. \cite{Maldacena:2002vr,Chen:2006xjb}. In wht follows, we shall utilize these values to extract the contribution from the one-loop effect. In the above-mentioned expressions, all the terms except the last one contributed as ${\cal O}(\epsilon^{2})$ and ${\cal O}(\epsilon^{3})$ both in SR and USR phases. On the other hand, since the last term contains the conformal time derivative of the factor $\eta/c^2_s$, the constraint regarding the sharp transition from the SR to USR phase has to be implemented on this particular contribution. Before the transition in the SR phase, contribution of the last term is insignificant. However, as the sharp transition is switched on, its immediate effect can be clearly seen in the USR phase, because its contribution is ${\cal O}(\epsilon)$. This is primarily due to the second slow roll parameter changing from $\eta\sim 0$ (in the SR phase) to $\eta\sim -6$ (in the USR phase) along with having a dynamical feature in the effective sound speed parameter $c_s$.

We are now explicitly computing the contribution of the third-order expanded EFT action's last term to the power spectrum of the scalar mode during PBH creation. Here, the last term of Eqn (\ref{s3}) is dominating over the other contributions in the USR phase. For this reason, in the one-loop result, the primordial scalar power spectrum contribution from this term is dominant over all the other terms. Such a dominant contribution is the only option which can accommodate large amplitude fluctuation in the USR regime, particularly during the formation of PBHs. However, for the sake of completeness during our computation, we have not neglected any of them though other terms are small during USR. Our derived result will capture both small and large contributions together in the final expression, which we will explicitly compute in this paper. To accommodate large amplitude fluctuation and explain the generation of PBHs, one can equivalently think in terms of an inflationary potential. By inserting a phenomenological bump or a dip in the structure of inflationary potential, one can mimic  this feature. Recently, in ref \cite{Mishra:2019pzq} with the help of a phenomenologically inserted Gaussian speed breaker, the authors have done the equivalent analysis within the framework of a canonical single field model of inflation to address this crucial issue at the tree level computation. But in this work our prime objective is to address the same issue from the perspective of EFT framework, where instead of a scalar field effective potential, we consider a more generalized version where the scalar perturbations are described in terms of Goldstone modes and at the starting point no scalar field kinetic term and its potential appear in the EFT action. All the features coming from bump or dip are clearly encoded in how we design the time dependent behaviour of the second slow-roll parameter $\eta$ and the effective sound speed $c_s$ during SR, USR, particularly to explain the SR to USR sharp transition. This also implies the fact that such insertions of bump or dip in the structure of the effective potential for inflation can be completely translated in the language of $\eta$ and $c_s$. As a result, within the EFT setup, no such information from the phenomenological structure of the inflation potential is actually required to describe the generation of PBHs from large amplitude fluctuations. There are only a few things we need to remember to visualize such mapping: (1) in our designed set up the PBHs formation scale $k_{\rm PBH}$ is the sharp transition scale $k_s$ from SR to USR, which is the position of the starting point of the bump or dip in the equivalent description, (2) Peak amplitude, which is ${\cal O}(10^{-2})$ in our description, can be equivalently described by the height/depth of the bump/dip of the phenomenological potential; and (3) the end of inflation as well as the end of the USR phase $k_e$ mimics the role of the finishing point of the bump or dip. 

In order to achieve this goal, we further employ the well-known in-in formalism. The following is an example of the leading cubic self-interaction Hamiltonian which plays the central role in our computation to generate the large amplitude fluctuation necessarily required for the formation of PBHs:
\bea H_{\rm int}(\tau)=-\frac{M^2_{ pl}}{2}\int d^3x\;  a^2\frac{1}{c^2_s}\epsilon\partial_{\tau}\left(\frac{\eta}{c^2_s}\right)\zeta^{'}\zeta^{2}.\eea
To implement the feature of the sharp transition, we need to design the behaviour of the effective sound speed, $c_s$, and the second slow roll parameter, $\eta$. In the SR and USR regimes, $\partial_{\tau}\left(\eta/c^2_s\right)\approx 0$ and this approximation fails at the sharp transition points, $\tau=\tau_s$ and $\tau=\tau_e$.  The only noteworthy contributions will thus occur at $\tau=\tau_s$ and $\tau=\tau_e$,  instead of the contribution from the complete time interval $-\infty<\tau<0$. During this construction, we also assume that at these transition points, the effective sound speed is approximately the same, i.e. $c(\tau_s)\approx c(\tau_e)=\tilde{c}_s\neq c_s$, where $c_s$ is the magnitude of the effective sound speed at the pivot scale. For this reason, at 
$\tau=\tau_s$ and $\tau=\tau_e$, we have, $\partial_{\tau}\left(\eta/c^2_s\right)\approx -\Delta\eta(\tau_s)/\tilde{c}^2_s$ and $\partial_{\tau}\left(\eta/c^2_s\right)\approx \Delta\eta(\tau_e)/\tilde{c}^2_s$, respectively. In this description, the effective sound speed at the transition scales has the structure, $\tilde{c}_s=1\pm \delta$, where $\delta$ is a fine-tuning factor. Consequently, due to having a non-uniform behaviour of the sound speed, at the sharp transition scales, one can accommodate the enhancement in the factor $\partial_{\tau}\left(\eta/c^2_s\right)$ which is necessarily required to provide an external kick to amplify the scalar perturbations to generate correct amplitude for the production of PBHs.

The tree level contribution to the total power spectrum from scalar perturbation considering both the SR and USR regime together can be further written as:
\begin{widetext}
    \bea &&\bigg[\Delta^{2}_{\zeta, {\bf EFT}}(p)\bigg]_{\bf Tree}=\bigg[\Delta^{2}_{\zeta,{\bf Tree}}(p)\bigg]_{\bf SR}+\bigg[\Delta^{2}_{\zeta,{\bf Tree}}(p)\bigg]_{\bf USR}=\bigg[\Delta^{2}_{\zeta,{\bf Tree}}(p)\bigg]_{\bf SR}\times\Bigg\{1+\Theta(p-k_s)\left(\frac{k_e}{k_s}\right)^6|\alpha_{\bf k}-\beta_{\bf k}|^2\bigg\},\quad\quad \eea
\end{widetext}

where $\Theta(p-k_s)$ is the Heaviside theta function inserted to implement the sharp transition from SR to USR regime. Here the scalar power spectrum's slow-roll (SR) contribution can be written as:
\bea &&\bigg[\Delta^{2}_{\zeta,{\bf Tree}}(p)\bigg]_{\bf SR}=\left(\frac{H^{2}}{8\pi^{2}M^{2}_{ pl}\epsilon c_s}\right)_{*}\left(1+\left(\frac{p}{k_s}\right)^2\right),\quad\quad\eea
where in the SR regime $p<k_s$ restriction is maintained. Here the term appearing at the parenthesis bracket is the amplitude of the scalar power spectrum in the SR region which is evaluated at the pivot scale $k_*=0.02{\rm Mpc}^{-1}$.

Further, in the presence of the previously mentioned third-order action for the comoving curvature perturbation and considering all the small and large contributions, we use the well-known in-in formalism in the present context along with all possible physically viable Wick contraction to generate one-loop Feynman diagrams. This further gives rise to the following simplified expressions for the one-loop contributions appearing from both the SR and USR regime:
\begin{widetext}
    \bea &&\bigg[\Delta^{2}_{\zeta, {\bf One-loop}}(p)\bigg]_{\bf SR}=\bigg[\Delta^{2}_{\zeta, {\bf Tree}}(p)\bigg]^2_{\bf SR}\times \Bigg(1-\frac{2}{15\pi^2}\frac{1}{c^2_{s}p^2_*}\left(1-\frac{1}{c^2_{s}}\right)\epsilon\Bigg)\Bigg(c_{\bf SR}-\frac{4}{3}{\cal I}_{\bf SR}(\tau_s)\Bigg),\\
    &&\bigg[\Delta^{2}_{\zeta, {\bf One-loop}}(p)\bigg]_{\bf USR}=\frac{1}{4}\bigg[\Delta^{2}_{\zeta, {\bf Tree}}(p)\bigg]^2_{\bf SR}\times \Bigg\{\Bigg(\frac{\left(\Delta\eta(\tau_e)\right)^2}{\tilde{c}^8_s}{\cal I}_{\bf USR}(\tau_e)-\frac{\left(\Delta\eta(\tau_s)\right)^2}{\tilde{c}^8_s}{\cal I}_{\bf USR}(\tau_s)\Bigg)-c_{\bf USR}\Bigg\},\quad\quad\quad\eea
\end{widetext}
In the above-mentioned one-loop contributions written for SR and USR region, the newly introduced factors ${\cal I}_{\bf SR}(\tau_s)$, ${\cal I}_{\bf USR}(\tau_s)$ and ${\cal I}_{\bf USR}(\tau_e)$ represents the contributions from the loop integrals in the region $\tau\leq \tau_s$, at the sharp transition scales, $\tau=\tau_s$ and $\tau=\tau_e$ respectively.  In the above-mentioned result the two factors $c_{\bf SR}$ and $c_{\bf USR}$ represent the regularization scheme dependent parameters which one needs to fix during performing the renormalization in the present context of the discussion.These factors can be computed in the late time scale, $\tau\rightarrow 0$ in the presence of UV and IR regulators of the underlying theory as:
\begin{widetext}
    \bea &&{\cal I}_{\bf SR}(\tau_s):=\lim_{\tau\rightarrow 0}\int^{k_s}_{p_*}\frac{dk}{k}\;\left(1+k^2c^2_s\tau^2\right)\approx\ln\left(\frac{k_s}{p_*}\right),\\
    &&{\cal I}_{\bf USR}(\tau_s):=\lim_{\tau\rightarrow 0}\int^{k_e}_{k_s}\frac{dk}{k}\;\bigg|\alpha_{\bf k}\left(1+ikc_s\tau\right)e^{-ikc_s\tau}-\beta_{\bf k}\left(1-ikc_s\tau\right)e^{ikc_s\tau}\bigg|^2\approx\ln\left(\frac{k_e}{k_s}\right),\\
    &&{\cal I}_{\bf USR}(\tau_e):=\left(\frac{k_e}{k_s}\right)^6\lim_{\tau\rightarrow 0}\int^{k_e}_{k_s}\frac{dk}{k}\;\bigg|\alpha_{\bf k}\left(1+ikc_s\tau\right)e^{-ikc_s\tau}-\beta_{\bf k}\left(1-ikc_s\tau\right)e^{ikc_s\tau}\bigg|^2\approx\left(\frac{k_e}{k_s}\right)^6{\cal I}_{\bf USR}(\tau_s).\eea
\end{widetext}
In the above-mentioned loop integrals, the upper limit corresponds to the UV cut-off and the lower one represents the IR cut-off of the underlying EFT setup in the corresponding SR and USR regime and the derived results represent the cut-off regularized contributions. Further, it is important to note that due to  the late time limit, $\tau\rightarrow 0$, one can clearly observe that the above-mentioned results are completely free from  quadratic UV divergences and can be expressed only in terms of the IR divergent logarithmic dependent factors in the final results. In the language of Quantum Field Theory of de Sitter space, such a limit actually serves the purpose of renormalization through which one can easily get rid of the quadratic divergences. These results can be rigorously obtained by applying an equivalent approach, known as wave function renormalization or adiabatic renormalization scheme which we have recently discussed in ref. \cite{Choudhury:2023vuj}. 

However, apart from the complete removal of the quadratic UV divergence contribution in the final result, IR logarithmic divergences appear which we need to soften by implementing the power spectrum renormalization method. In this paper, we have performed our analysis in a completely model-independent fashion in the language of EFT, see also refs. \cite{Choudhury:2023rks} where we did the rigorous analysis with  a sharp transition from SR to USR region for canonical and $P(X,\phi)$ single field slow-roll models of inflation. Additionally, it is important to note that during performing the one-loop momentum integrals we have introduced constant factors, $c_{\bf SR}$ and $c_{\bf USR}$, which are represent the renormalization scheme dependent parameters for the SR and USR regions respectively.
he total contribution to the one-loop corrected power spectrum for the scalar perturbation can be further written in a renormalization scheme independent fashion as:
    \bea &&\Delta^{2}_{\zeta, {\bf EFT}}(p)
=\bigg[\Delta^{2}_{\zeta,{\bf Tree}}(p)\bigg]_{\bf SR}\Bigg\{1+U+V\Bigg\},\quad\quad\eea
where  $U$ and $V$ are defined as:
\bea 
&&U=-\frac{4}{3}\bigg[\Delta^{2}_{\zeta,{\bf Tree}}(p)\bigg]_{\bf SR}\nonumber\\
&&\times\Bigg(1-\frac{2}{15\pi^2}\frac{1}{c^2_{s}p^2_*}\left(1-\frac{1}{c^2_{s}}\right)\epsilon\Bigg)
\ln \left(\frac{k_s}{p_*}\right),\eea
\bea
&&V=\frac{1}{4}\bigg[\Delta^{2}_{\zeta,{\bf Tree}}(p)\bigg]_{\bf SR}\nonumber\\
&&\times \Bigg(\frac{\left(\Delta\eta(\tau_e)\right)^2}{\tilde{c}^8_s} \left(\frac{k_e}{k_s}\right)^{6}-\frac{\left(\Delta\eta(\tau_s)\right)^2}{\tilde{c}^8_s}\Bigg)\ln \left(\frac{k_e}{k_s}\right).\quad\eea


The wave numbers $k_e$ (UV cut-off) and $k_ s$ (IR cut-off) that correspond to the time scales, $\tau_e$, and $\tau_s$ should also be taken into consideration.

Now, in order to soften the impacts of one-loop logarithmic divergences from SR as well as the USR region, we additionally define the renormalized power spectrum for the scalar perturbation for the prescribed EFT setup as, 
\be \overline{\Delta^{2}_{\zeta,{\bf EFT}}(p)}={\cal Z}_{\zeta,{\bf EFT}}\Delta^{2}_{\zeta, {\bf EFT}}(p),\ee
where the explicit renormalization condition determines the renormalization factor, often known as the counter-term,  which is represented by the equation ${\cal Z}_{\zeta,{\bf EFT}}$.  In the current framework,  the pivot scale $p *$,  which is supplied by the formula,  is fixed as the equivalent renormalization condition, \be \label{bcd}\overline{\Delta^{2}_{\zeta,{\bf EFT}}(p_*)}=\bigg[\Delta^{2}_{\zeta,{\bf Tree}}(p_*)\bigg]_{\bf SR},\ee
using which the counter term,  ${\cal Z}_{\zeta,{\bf EFT}}$ can be computed as, 
\be {\cal Z}_{\zeta,{\bf EFT}}\approx\left(1-U_*-V_*\right).\ee
Then using this counter term the corresponding one-loop corrected renormalized power spectrum for the scalar modes can be expressed as:
\bea  \overline{\Delta^{2}_{\zeta,{\bf EFT}}(p)}&=&\bigg[\Delta^{2}_{\zeta,{\bf Tree}}(p)\bigg]_{\bf SR}\Bigg\{1+{\cal Q}_{\bf EFT}\Bigg\},\eea
where we define ${\cal Q}_{\bf EFT}$ as:
\bea 
{\cal Q}_{\bf EFT}&=&-\frac{\bigg[\Delta^{2}_{\zeta,{\bf Tree}}(p)\bigg]_{\bf SR}}{\bigg[\Delta^{2}_{\zeta,{\bf Tree}}(p_*)\bigg]_{\bf SR}}\Bigg\{U^{2}_*+V^{2}_*+\cdots\Bigg\}.\eea
Before we go into the technical parts of the calculation, let us state unequivocally that the current calculation is not dependent on the explicit mathematical structure of the contributions from one-loop correction. To conduct the resummation and finally deliver a finite output, the perturbative approximations are maintained throughout. We provide the results in a way that demonstrates how well this strategy works within the present framework.

We finally briefly describe the Dynamical Renormalization Group (DRG) approach \cite{Boyanovsky:1998aa,Boyanovsky:2001ty,Boyanovsky:2003ui,Burgess:2009bs,Dias:2012qy,Chen:2016nrs,Baumann:2019ghk,Burgess:2009bs,Chaykov:2022zro,Chaykov:2022pwd,Choudhury:2023vuj}, which enables us to resum over all of the contributions that are logarithmically divergent. Technically speaking, this is feasible as long as the related resummation infinite series is strictly convergent at late time scales.   Actually, every term in this series is a direct outcome of the perturbative expansion in every feasible loop order. In summary, DRG is seen as the natural technique via which the validity of secular time-dependent contributions will be justified in the cosmological perturbative expansion. Using this technique one can absorb the secular contributions to the running coupling parameters of the theory, which is further referred to as the Renormalization Group resummation method in the present context of discussion. Within the framework of Cosmology, such running coupling parameters can be easily identified with the spectral tilt, running, and running of the running of the spectral tilt of the scalar power spectrum, which in the language of Quantum Field Theory are physically interpreted as the Beta functions. This has a deeper connection with the previously discussed renormalization procedure discussed in this paper. One has to renormalize and resum the scalar power spectrum in such a fashion that at the CMB pivot scale all of these beta functions computed from the SR phase and USR phase give the same result. This is very crucial information and in the present computation, one can think of this as a matching condition that needs to be maintained throughout the computation. One can easily verify the viability of this statement from Eq. (\ref{bcd}), which further helps us fix the structure of the counterterm for the power spectrum renormalization, hence the DRG resumed result of the power spectrum.

Using the DRG method, the final form of the resummed dimensionless power spectrum can be expressed as follows:
\bea &&\overline{\overline{\Delta^{2}_{\zeta,{\bf EFT}}(p)}}=\bigg[\Delta^{2}_{\zeta,{\bf Tree}}(p)\bigg]_{\bf SR}\exp\bigg({\cal Q}_{\bf EFT}\bigg)\nonumber\\
&&\quad\quad\quad\quad\quad\quad \times\bigg\{1+\bigg[\Delta^{2}_{\zeta,{\bf Tree}}(p_*)\bigg]_{\bf SR}\bigg\}.\eea
Here, $|{\cal Q}_{\bf EFT}|\ll 1$, the rigorous convergence criteria for the DRG resummed infinite series as seen in the elucidation,  is met.  In the above-mentioned result, the first term in the parenthesis bracket corresponds to the leading order result which appears as a direct outcome of the DRG exponentiation. On the other hand, the second term mimics the role of a correction term which in the language of perturbation theory can be translated as the sub-leading contribution activated only at the pivot scale of the computation. Thus, when all the possible terms from the previously mentioned secular components are combined together, it really finally depicts the behaviour on a large scale, which from the perspective of cosmological perturbation theory is extremely important information. 
We found that after DRG resummation the overall signature in front of the factor ${\cal Q}_{\bf EFT}$ is negative which appears in the exponent, which further implies a sharp fall of the spectrum at the end of USR phase and at the end of inflation. This is a unique finding of this present work which is directly reflected in the behaviour of the spectrum and from this obtained result it is obvious that the resumed spectrum severely constrains the appearance of PBH formation scale (which coincides with the sharp transition scale) and hence constraint the generated PBH mass to be small. Proper estimations are provided in the later half of this paper (see refs. \cite{Choudhury:2023vuj,Choudhury:2023rks} where we have performed the DRG resummation in the context of canonical and $P(X,\phi)$ models of single-field inflation.)

Our result is depicted in fig.(\ref{N1}-\ref{N3}), we have shown the behaviour of the dimensionless power spectrum for scalar modes with respect to the number of e-foldings for the effective sound speed $c_s=0.6$ with $M^4_2/\dot{H}M^2_p\sim -0.89$ (non-canonical and causal), $c_s=1$ with $M^4_2/\dot{H}M^2_p\sim 0$  (canonical and causal) and $c_s=1.17$ with $M^4_2/\dot{H}M^2_p\sim 0.13$ (non-canonical and a-causal) respectively. Here we fix $k_s=10^{21}\;{\rm Mpc}^{-1}$, \textcolor{red}{$k_e=10^{22}\;{\rm Mpc}^{-1}$}, $p_*=0.02\;{\rm Mpc}^{-1}$, $c_{\bf SR}=0$, $0\lesssim M^4_2/\dot{H}M^2_p\lesssim 0.13$ for our analysis. From this plot we have found that, 
$\Delta {\cal N}_{\rm USR}=\ln(k_e/k_s)\approx\ln(10)\approx 2,$ 
which implies around $2$ e-folds are allowed in the USR period for the PBH formation. 
We discovered that among the canonical ($c_s=1$, see fig. \ref{N2}) and a-causal ($c_s>1$, see fig. \ref{N3}) frameworks, $c_s>1$ is preferred because the peak of the spectrum precisely reaches the value $10^{-2}$ for $c_s=1.17$ (see fig. \ref{N3}), which is the desired amplitude of the spectrum for PBH formation\footnote{Bringing in the non-canonical features with $c_s<1$ worsens the chances for PBHs formation.}. 

If we  further increase the sound speed, $c_s>1.17$,\footnote{We found that for $c_s=3/2$ with the coupling $M^4_2/\dot{H}M^2_p\sim 0.28$, the amplitude of the scalar power spectrum reaches at the maximum value ${\cal O}(1)$, where the perturbation theory strictly break down. } the perturbation would break down during PBH production. The field excursion also suggests that the EFT prescription is valid during the USR period when the PBH formation takes place and one can consider sub-Planckian EFT of single field inflation for PBH formation.

Since we are working within the framework of EFT, we will notice a few changes due to the involvement of the effective sound speed $c_{s}$ in the formula for the mass of PBH. Following the derivation in ref.\cite{Mishra:2019pzq} for the PBH mass, we arrive at a similar result but with the mentioned changes as shown below. We start with the fact that due to the EFT framework, we have the modification $c_{s}(\tau)k_{\rm s,e} = (aH)_{\rm s,e}$ for time $\tau = \tau_{s}\;{\rm and}\;\tau = \tau_{e}$ where we have sharp transitions. This means that the horizon crossing condition gets modified due to having effective sound speed. Now, as discussed earlier, the sound speed parameter has the parametrization, $c_{s}(\tau_{s}) = c_{s}(\tau_{e}) = \tilde{c}_{s} = 1\pm\delta$ where $\delta \ll 1$. Thus, we observe that $\tilde{c}_{s}k_{\rm s,e} = (aH)_{\rm s,e}.$ Now, for the pivot scale $k = p_{*}$ we also have $c_{s}(\tau_{*}) = c_{s,*} = c_{s}$, and hence we get $c_{s}p_{*} = (aH)_{*}$. With these useful facts in mind, we proceed toward the derivation for the PBH mass as follows.
The Hubble scale value when concerned with the radiation-dominated (RD) era follows the relation:
\bea \label{Hpbh}
H^{2} &=& \frac{\rho_{r}}{3M_{p}^{2}} = \Omega^{0}_{r}H_{0}^{2}\frac{\rho_{r}}{\rho^{0}_{r}}\nonumber\\
&=& \Omega^{0}_{r}h^{2}\times\left(\frac{100 {\rm km}}{\rm s\;Mpc}\right)^{2}\left(\frac{g_{*}}{g^{0}_{*}}\right)\left(\frac{T}{T^{0}}\right)^{4},
\eea
where the relation between energy densities and temperature is used and the notation $``0"$ denotes the respective quantity evaluated at the present time scale. Now, using the conservation of entropy and the assumption that the effective number of relative degrees of freedom of the energy and entropy densities, evaluated during the Radiation Dominated (RD) era, are almost equal $g_{*} \sim g_{*,s}$, we get the following expression for the Hubble scale in the RD era:
\bea
H^{2} = \Omega^{0}_{r}h^{2}\left(\frac{100 {\rm km}}{\rm s\;Mpc}\right)^{2}\left(\frac{g_{*}}{g_{0*}}\right)^{-1/3}\left(\frac{g^{0}_{*,s}}{g^{0}_{*}}\right)^{4/3}(1+z)^{4},\quad\quad\eea
where the redshift factor comes in the present computation due to entropy conservation. Now the mass of the PBHs generated due to the large fluctuations entering during the RD era depends on the respective Horizon mass as follows:
\bea M_{\rm PBH} = \gamma M_{H} = \gamma\left(\frac{4\pi M_{p}^{2}}{H}\right), \eea
where $\gamma(\sim 0.2)$ is the critical collapse factor. From the expression for the Hubble scale derived above and the values for the degrees of freedom $g^{0}_{*}=3.38$, $g^{0}_{*,s}=3.94$, and $\Omega^{0}_{r}h^{2}=4.18\times10^{-5}$, we get the following relation for the mass of PBH:
\bea \label{zpbh}
\frac{M_{\rm PBH}}{M_{\odot}}&=&1.55\times10^{24}\nonumber\\
&\times&\left(\frac{\gamma}{0.2}\right)\left(\frac{g_{*}}{106.75}\right)^{1/6}(1+z)^{-1/2}, \eea

Further utilizing the fact that the Hubble scale during inflation is almost a constant, we use the known relation $(aH)_{*} = c_{s}p_{*}$ for the wavenumber $p_{*}$ which at the time exits the Hubble radius and, as a result of this assumption, for the wavenumber which exits at the time of the PBH formation we also have the relation $(aH)_{\rm PBH} \sim a_{\rm PBH}H_{*}$. From this, we get a result for the amount of expansion in terms of the number of e-foldings:
\bea \Delta N = \ln\left({\frac{a_{\rm PBH}}{a_{*}}}\right) = \ln\left({\frac{(aH)_{\rm PBH}}{(aH)_{*}}}\right) = \ln\left({\frac{\tilde{c}_{s}k_{\rm PBH}}{c_{s}p_{*}}}
\right),\quad\quad\eea
where we have used the fact that PBH formation occurs at the scale $k_{\rm PBH} = k_{s}$ and also have used the effective sound speed at that time $\tau=\tau_{s}$ is $\tilde{c}_{s}$. Using $a_{\rm PBH} = 1/(1+z)$, the second equality in the above equation, when combined with the use of Hubble scale at PBH formation from eqn.(\ref{Hpbh}), and the relation between redshift and mass from eqn. (\ref{zpbh}), gives us the relation:
\bea
&&\Delta N = 17.33 + \frac{1}{2}\ln\left({\frac{\gamma}{0.2}}\right)\nonumber\\
&&\quad\quad\quad\quad-\frac{1}{12}\ln\left({\frac{g_{*}}{106.75}}\right)-\frac{1}{2}\ln\left({\frac{M_{\rm PBH}}{M_{\odot}}}\right). \eea
Exponentiation of this above-mentioned expression gives us the desired result:
\bea
\left(\frac{M_{\rm PBH}}{M_{\odot}}\right)_{c_{s}}&=&1.13\times10^{15}\nonumber\\
&\times&\left(\frac{\gamma}{0.2}\right)\left(\frac{g_{*}}{106.75}\right)^{-1/6}\left(\frac{c_{s}p_{*}}{\tilde{c}_{s}k_{s}}\right)^{2},
\eea
where the final mass is quantified in terms of the formation scale $k_{\rm PBH} = k_{s}$, which coincides with the sharp transition scale in our present computation. Here $M_{\odot}\sim 2\times 10^{30}{\rm kg}$ is the solar mass, $\gamma\sim 0.2$ is the efficiency factor and $g_*$ is the relativistic d.o.f.. Also, the pivot scale is fixed at, $p_*=0.02\;{\rm Mpc}^{-1}$. We know that from the previously mentioned behaviour of the sound seed, $c_s\neq \tilde{c}_s$ and both the possibilities $\tilde{c}_s>c_s$ and $\tilde{c}_s<c_s$ can be accommodated. Consequently, the above-mentioned formula can be further simplified as:
\bea &&\left(\frac{M_{\rm PBH}}{M_{\odot}}\right)_{c_s}=1.13\times 10^{15}\times\bigg(\frac{\gamma}{0.2}\bigg)\bigg(\frac{g_*}{106.75}\bigg)^{-1/6}\nonumber\\
&&\quad\quad\quad\quad\quad\quad\quad\times\left(\frac{p_*}{k_s}\right)^{2}\times c^2_s(1\mp 2\delta)\nonumber\\
&&\quad\quad\quad\quad\quad\;\approx \left(\frac{M_{\rm PBH}}{M_{\odot}}\right)_{c_s=1}\times c^{2}_s,\quad\eea
where the contribution from the small fine-tuning factor $\delta$ is neglected for the sake of simplicity. It is important to note that, for the canonical single field $c_s=1$ case we have \cite{Choudhury:2023vuj}:
\bea &&\left(\frac{M_{\rm PBH}}{M_{\odot}}\right)_{c_s=1}=1.13\times 10^{15}\nonumber\\
&&\quad\quad\quad\quad\quad\quad\quad\quad\times\bigg(\frac{\gamma}{0.2}\bigg)\bigg(\frac{g_*}{106.75}\bigg)^{-1/6}\left(\frac{p_*}{k_s}\right)^{2}.\quad\quad\eea
Here the prime motivation is to express the PBH mass for the EFT setup in terms of the contribution obtained from the canonical single-field ($c_s=1$) setup. It further helps us to clearly visualize that due to having the EFT setup only modification appears in this expression in terms of the effective sound speed, and within a very small preferred window the contribution is small. However, we have explicitly pointed out this contribution to quantify the change in the presence of the EFT setup.

For further numerical estimation, we fix $\gamma\sim 0.2$,  $g_*\sim 106.75$, $p_*=0.02\;{\rm Mpc}^{-1}$, and $k_s=10^{21}{\rm Mpc}^{-1}$:
\bea \left(\frac{M_{\rm PBH}}{M_{\odot}}\right)_{c_s=1}=4.52\times {10}^{-31},\eea 
which finally gives the following result:
\bea &&\left(\frac{M_{\rm PBH}}{M_{\odot}}\right)_{c_s}\approx 4.52\times {10}^{-31}\times c^{2}_s.\quad\eea

In ref.\cite{Sasaki:2018dmp,Kawasaki:2016pql}, instead of using pivot scale, the authors have used the wavenumber at radiation-matter equality $k_{\rm eq}$ for the PBH mass estimation. Also, this computation was performed for the canonical single field slow-roll inflation model where $c_{s}=1$. Since we are doing the analysis in the presence of EFT, it is expected to have a dependence on the sound speed throughout the derivation in the subsequent steps, as already discussed above. By following the normalization scale of $k_{\rm eq}$ as in ref.\cite{Sasaki:2018dmp,Kawasaki:2016pql}, if we repeat the analysis as done in the above-mentioned steps, then we can write down the final expression as:
\bea
\frac{M_{\rm PBH}}{M_{\odot}}&=& 3.6\left(\frac{\gamma}{0.2}\right)\left(\frac{g_{*}}{106.75}\right)^{-1/6}\left(\frac{\tilde{c}_{s}k_{\rm PBH}}{c_{s}10^{6}{\rm Mpc}^{-1}}\right)^{-2}
\eea
where the scale dependent effective sound speed $\tilde{c}_{s}\;{\rm and}\;c_{s}$ are included. Now, writing this expression using the scale $k_{\rm eq}$, and its corresponding sound speed denoted as $\tilde{\tilde{c}}_{s}$, we get the following:
\bea
\frac{M_{\rm PBH}}{M_{\odot}}&=& 3.6\left(\frac{\gamma}{0.2}\right)\left(\frac{g_{*}}{106.75}\right)^{-1/6}\nonumber\\
&\times&\left(\frac{\tilde{c}_{s}k_{\rm PBH}}{\tilde{\tilde{c}}_{s}k_{\rm eq}}\right)^{-2}\left(\frac{\tilde{\tilde{c}}_{s}k_{\rm eq}}{10^{6}{\rm Mpc}^{-1}}\right)^{-2} \eea
substituting the values of the parameters $k_{\rm eq} = 0.07\Omega^{0}_{m}h^{2}{\rm Mpc}^{-1}$, where $h\sim 0.674$ and $\Omega^{0}_{m}\sim 0.315$ gives us the relation:
\bea
\frac{M_{\rm PBH}}{M_{\odot}}&=& 0.36\times10^{15}\times\left(\frac{\gamma}{0.2}\right)\nonumber\\
&\times&\left(\frac{g_{*}}{106.75}\right)^{-1/6}
\left(\frac{\tilde{c}_{s}k_{\rm PBH}}{\tilde{\tilde{c}}_{s}k_{\rm eq}}\right)^{-2} \eea
from this result, we make use of the fact that since the wavenumber at the radiation-matter equality scale is very near to the wavenumber at pivot scale, the effective sound speed remains almost constant within this short interval and as a result, we can assume $\tilde{\tilde{c}}_{s} \simeq c_{s}$ in this case. Hence, when we plug in the values for $\gamma \sim 0.2$, $g_{*}=106.75$, and $k_{\rm PBH}=k_{s}\sim 10^{21}{\rm Mpc}^{-1}$ and use the value for $\tilde{c}_{s}=1+\delta$, such that $\delta \ll 1$, we arrive at the following result for the expression for  PBH mass:
\bea
\left(\frac{M_{\rm PBH}}{M_{\odot}}\right)_{c_{s}} \approx 10^{-31}\times c^{2}_{s}.
\eea

From the above results, we make an assertion about the allowed PBH mass produced from the framework of EFT of single-field inflation, by working out a calculation of the fractional abundance of the PBH produced when the contribution from the one-loop corrected power spectrum is taken into account. This requires knowing about the mass fraction of the PBHs at formation which is given as:
\bea f_{\rm PBH}&=&1.68\times 10^{8}\times\left(\frac{\gamma}{0.2}\right)^{1/2}\nonumber\\
&\times&\left(\frac{g_{*}}{106.75}\right)^{-1/4}\left(\frac{M_{\rm PBH}}{M_{\odot}}\right)^{-1/2}\beta(M_{\rm PBH}).\quad \eea
Now, due to the requirement of an over-dense region to allow for the formation of PBH as the large density perturbations enter the Horizon at a certain scale after inflation, if the density contrast increases to a certain threshold value $\delta_{\rm th} > \delta_c=1/3$, then the probability of that event calculated using the Press-Schechter formalism gives:
\bea
\beta(M_{\rm PBH}) 
&\approx& \gamma\left(\frac{\sigma_{M_{\rm PBH}}}{\sqrt{2\pi}\delta_{\rm th}}\right)\exp{\left(-\frac{\delta_{\rm th}^{2}}{2\sigma_{M_{\rm PBH}}^{2}}\right)}.\eea
Here the quantity $\delta_{\rm th}$ is coarse-grained over the scale of PBH formation $R = 1/(\tilde{c}_{s}k_{\rm PBH}) = 1/(aH)_{\rm PBH}$. Also $\sigma_{M_{\rm PBH}}$ represents the variance of the coarse-grained density fluctuations at the mass scale $M_{\rm PBH}$. The appearance of the effective sound speed parameter is an outcome of present EFT framework, where $c_{s}=1$ reduces to the canonical single-field inflation  \cite{Mishra:2019pzq,Sasaki:2018dmp,Kawasaki:2016pql}.

The smoothening of these density fluctuations over the scale $R$ is done through the use of a Gaussian window function $W(p,R)=\displaystyle{\exp{(-p^{2}R^{2}/2)}}$, and the corresponding variance is computed as:
\bea
\sigma_{M_{\rm PBH}}^{2}&=&\int^{\infty}_{0} d\ln{p}\;\Delta^{2}_{\delta}(p)W^{2}(p,R), 
\eea
where the power spectrum of the density contrast can be expressed in the RD era as:
\bea \label{power} \Delta^{2}_{\delta}(p)&=&\frac{16}{81}\left(\frac{p}{\tilde{c}_{s}k_{\rm PBH}}\right)^{4}\overline{\overline{\Delta^{2}_{\zeta,{\bf EFT}}(p)}}. \eea
     \begin{figure}[htb!]
    	\centering
 \subfigure[For $c_s=1$  with $M^4_2/\dot{H}M^2_p\sim -0.89$ (non-canonical and causal).]{
      	\includegraphics[width=8.7cm,height=6.3cm] {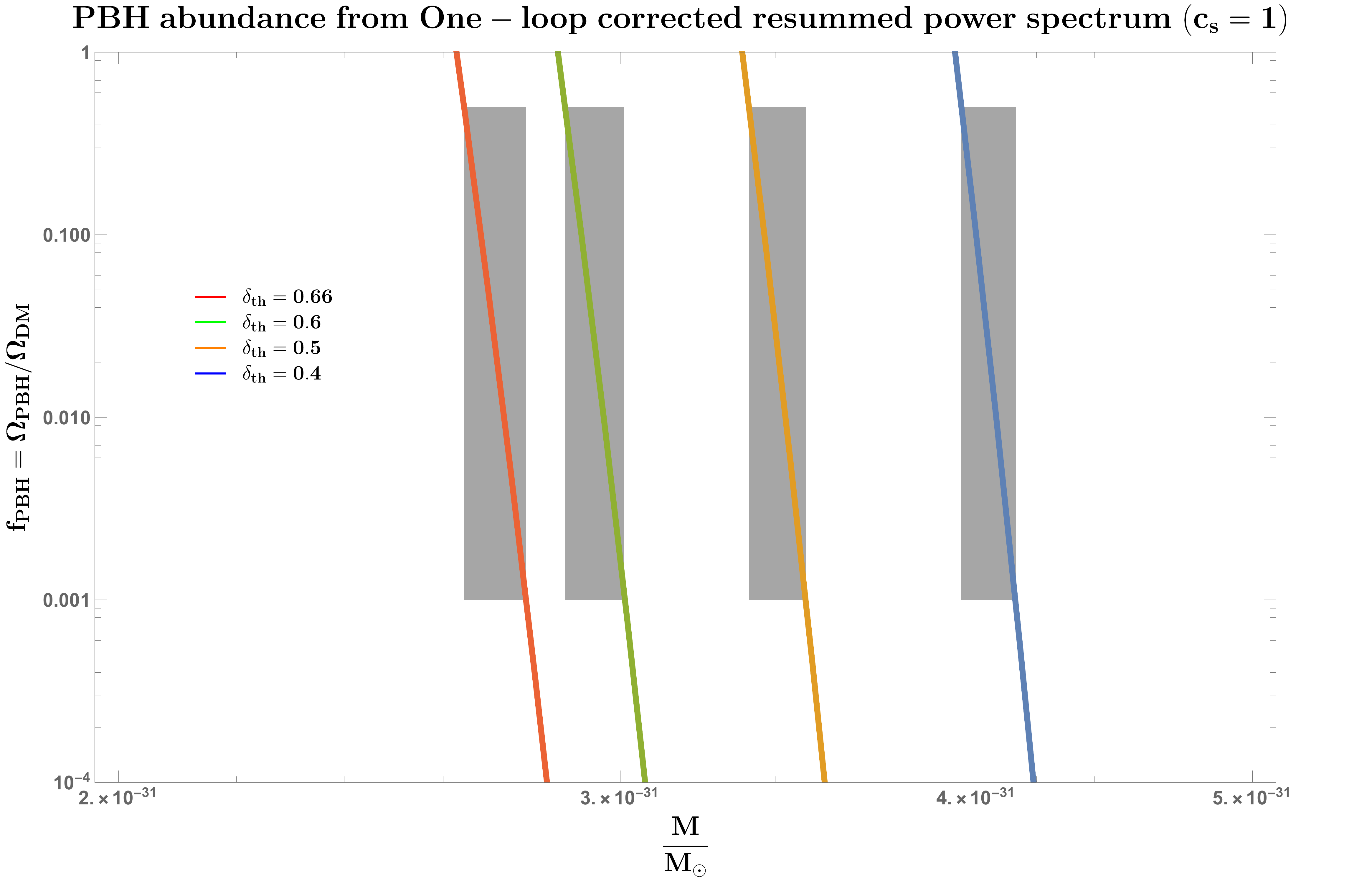}
        \label{ab1}
    }
    \subfigure[For $c_s=1.17(>1)$  with $M^4_2/\dot{H}M^2_p\sim 0.13$ (non-canonical and a-causal).]{
       \includegraphics[width=8.7cm,height=6.3cm] {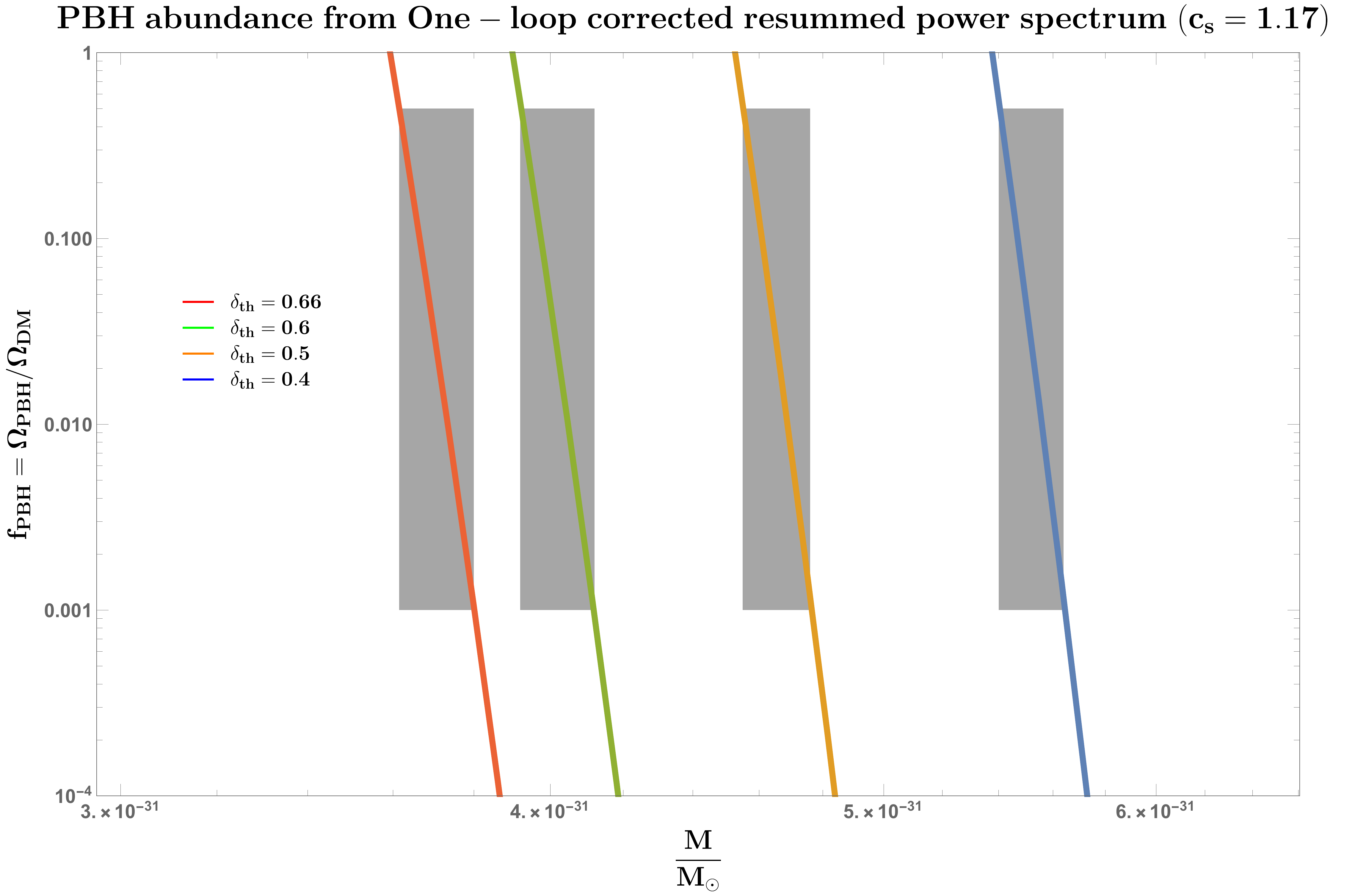}
        \label{ab2}
       }
    	\caption[Optional caption for list of figures]{Behaviour of the abundance of PBHs when plotted against their mass. The effective sound speed is taken to be as $c_{s}=1\;{\rm and}\;1.17$. The gray shaded region suggests the allowed mass region of the PBH produced for the corresponding density contrast threshold where the abundance falls in between $0.001 \leq f \leq 0.5$. Based on the constraint for the threshold of collapse necessary to neglect the non-linearities in the density contrast, the interval $2/5 \leq \delta_{\rm th} \leq 2/3$ is considered and the range of PBH mass, for both the cases of $c_{s}$ values, lies in between $10^{2}{\rm gm} < M_{\rm PBH} \lesssim 10^{3}{\rm gm}$.} 
    	\label{abundance}
    \end{figure}

While calculating the PBH abundance produced from the present EFT framework, we must consider the value for the density contrast threshold within the interval $2/5 \leq \delta_{\rm th} \leq 2/3$, see \cite{Musco:2018rwt}, since this is the regime were linearities in the density contrast dominate. Recently, there have been important discussions regarding the case of non-linear features in the density contrast and its effects on the final threshold range, see refs.\cite{Musco:2020jjb, DeLuca:2023tun} for details. For the values lying in the regions $1/3 < \delta_{\rm th} < 2/5$ and $2/3 < \delta_{\rm th} < 1$, the non-linearities in the density contrast come into play and involves the study of non-Gaussian features \footnote{The underlying connecting relationship between the density contrast and the comoving curvature perturbation in the coordinate space is given by the following expression \cite{Harada:2015yda}:
\bea \delta(\mathbf{x}, t) = -\frac{4}{9}\frac{1}{a^{2}H^{2}}\exp{(-2\zeta(\mathbf{x}))}\left(\nabla^{2}\zeta(\mathbf{x})+\frac{1}{2}\partial_{i}\zeta(\mathbf{x})\partial^{i}\zeta(\mathbf{x})\right),\quad
\eea
where from this expression, the leading order term in the quantity $\zeta(\mathbf{x})$ results in the expression:
\bea \delta(\mathbf{x}, t) = -\frac{4}{9a^{2}H^{2}}\nabla^{2}\zeta(\mathbf{x}). \eea This expression when transformed into the Fourier space results in the exact relation between the power spectrum for the density contrast and the curvature perturbation in eqn.(\ref{power}) within the EFT framework. The same sort of linear approximation is also taken initially when writing the scalar fluctuation component in the metric.}
which is beyond the scope of our present analysis. To this end, having the density contrast profile obey Gaussian statistics and with the Gaussian window function chosen as before, it is sufficient to smoothen the density contrast which helps us to calculate the PBH abundance factor. Ultimately, an intrinsic non-linear relation exists between the density contrast and the curvature perturbation, which when considered is able to provide a more robust  understanding of the PBH production related issues. The following refs.\cite{DeLuca:2019qsy, Young:2019yug}, have their discussions centered around this non-linear relation being present, where the authors find that it is  much more challenging to generate PBHs by relying on perturbation theory and require the use of different methods, like peak theory and other methods of statistics for the non-Gaussian threshold, to deal with these intrinsic non-Gaussian features and the resulting probability of PBH formation.

We now present the outcomes of the study in terms of the abundance plots when the effective sound speed is $c_{s} = 1$ and $c_s=1.17$ and from the representative plots we determine the allowed range of the resulting PBH mass. We find the allowed range of PBHs masses, namely, $10^{2} {\rm gm} \lesssim M_{\rm PBH}\lesssim 10^{3} {\rm gm}$ corresponding to  $2/5 \leq \delta_{\rm th} < 2/3$.


From the figures (\ref{ab1}-\ref{ab2}), we infer that there exists a range of masses of PBH generated within the framework of EFT of single-field inflation which accounts for a significant contribution to the dark matter density when measured today. This analysis is carried out based on the numerically allowed region for the density contrast, $\delta_{\rm th}$,  where the non-linearities present in this quantity can be ignored within the constraint $2/5 \lesssim \delta_{\rm th} \lesssim 2/3$. The mass values are highly sensistive to its particular threshold value as seen in the plots above from their shaded regions in each of the plots drawn for effective sound speeds $c_s=1$ and $c_s=1.17$ respectively. 
The analysed set of values for the threshold also has a clear advantage to maintain the perturbativity approximation in our computation. For values closer to the upper bound $\delta_{\rm th}\sim 2/3$, we found it leads to the generation of PBHs with $M_{\rm PBH} \sim 10^{2}{\rm gm}$ while for the values near the threshold lower bound $\delta_{\rm th}\sim 2/5$ gives us PBH with $M_{\rm PBH} \sim 10^{3}{\rm gm}$. 
It is challenging to produce PBH with $M_{\rm PBH} \lesssim 10^{2}{\rm gm}$ as it requires us to increase the threshold value beyond the mentioned upper bound, which brings in the non-linearities into the picture and even to consider the region $\delta_{\rm th} > 1$ which directly corresponds to the breakdown of perturbation theory and strictly not allowed in the present computation. Also, we observe that for the range of PBH masses shown in the plot the allowed band of values, for a specific threshold of density contrast, is very small when considering a significant abundance region which implies that there exists a limited range of PBH masses that can contribute to the dark matter density where the smaller mass PBHs are produced above a higher threshold value.

The non-Gaussianities in the density contrast are critical for the better understanding of the PBH production scenario and ultimately provide more realistic values for the PBH abundance from theory. We have recently made some progress in this direction, see refs.\cite{Choudhury:2023hvf, Choudhury:2023kdb}, after studying the non-Gaussianities in Galileon theory through the use of the in-in formalism. We have found that the large non-Gaussianities can be generated in a controlled fashion and within this theory, it is further possible to evade the \textit{no-go theorem}. The problem regarding the non-Gaussian features in the density contrast can also be tackled through more rigorous analysis\cite{DeLuca:2019qsy, Young:2019yug}; we defer these investigations  to our future work.

In conclusion, we have found that within the framework of EFT of single field inflation, the allowed span for PBH formation in terms of the number of e-foldings is approximately $2$, the estimated PBH mass is extremely small and lies within the interval $10^{2}{\rm gm}\lesssim M_{\rm PBH}\lesssim 10^{3} \rm gm$ for the threshold value within the interval $2/5 \leq \delta_{\rm th} \leq 2/3$ which leads to the PBH abundance in the physically significant region of $0.001 \lesssim f \lesssim 0.5.$. In this computation, the allowed window for the effective sound speed is given by, $1<c_s<1.17$, for the gravitational coupling ,$0\lesssim M^4_2/\dot{H}M^2_p\lesssim 0.13$. Additionally, we have seen from our analysis that both canonical and a-causal frameworks, ($c_s\geq 1$) are allowed, out of which the a-causal one is most favoured due to having maximum enhancement of the PBH power spectrum.  We should note that our findings are strictly valid for the sharp transition from SR to USR. In this case, the possibility of generating large masses of PBHs in models of single-field inflation (canonical and non-canonical) is ruled out.
\\

{\bf Acknowledgements:} We thank 
Sergey Ketov and M. R. Gangopadhyay for useful discussions. The work of MS is supported by Science and Engineering Research Board (SERB), DST, Government of India under the Grant Agreement number CRG/2022/004120 (Core Research Grant). MS is also partially supported by the Ministry of Education and Science of the Republic of Kazakhstan, Grant
No. 0118RK00935 and CAS President's International Fellowship Initiative (PIFI).


\end{document}